\documentclass[twocolumn,showpacs,preprintnumbers,aps]{revtex4}

\usepackage[dvips]{graphicx}
\usepackage{amsfonts}
\usepackage{dcolumn}
\usepackage{bm}

\newtheorem{prop}{Proposition}[section]
\begin{document}
\title{
Quantum machine language  and quantum computation with Josephson
junctions}

\author{K. Ch. Chatzisavvas}\email{kchatz@auth.gr}
\author{C. Daskaloyannis}\email{daskalo@auth.gr}
\author{C. P. Panos}\email{chpanos@auth.gr}
\affiliation{Physics Department,\\
             Aristotle University of Thessaloniki,\\
             54006 Thessaloniki, Greece}

\date{September, 2001}


\begin{abstract}

An implementation  method  of a  gate in a quantum computer is
studied in terms of a finite number of steps evolving in time
according to a finite number of basic Hamiltonians, which are
controlled by on-off switches.  As a working example, the case of
a particular implementation of the two qubit computer employing a
simple system of two coupled Josephson junctions is considered.
The binary values of the switches together with the time durations
of the steps  constitute the quantum machine language of the
system.
\end{abstract}

\maketitle

\section{Introduction}\label{sec:Introduction}

In classical computing the programming is based on commands
written in the machine language. Each command is translated into
manipulations of the considered device, obtained by electronic
switches.
 In quantum computation quantum mechanics
is employed to process information. Therefore the conception of
the quantum computer programming, the structure of the quantum
machine language and the commands  are expected to be quite
different from the classical case. Although there are differences
between classical and quantum computers, the programming in both
cases should be based on commands, and a part of these commands is
realized by quantum gates.
 Initially, one of the leading ideas
in quantum computation was the introduction of the notion of the
universal gate \cite{DeuPRSL85}. Given the notion of a universal
set of elementary gates, various physical implementations of a
quantum computer have been proposed \cite{BraLoFDP00}. Naturally,
in order for an implementation to qualify as a valid quantum
computer, all the set of elementary gates have to be implemented
by the proposed system. This property is related to the problem of
controllability of the quantum computer.The controllability of
quantum systems is an open problem under investigation
\cite{RaViMoKo00}. Recently, \cite{DiVincpre01, DiVincNa00}
attention was focused on the notion of encoded universality, which
is a different functional approach to the quantum computation.
Instead of forcing a physical system to act as a predetermined set
of universal gates, which will be connected by quantum
connections, the focus of research is proposed to be shifted to
the study of the intrinsic ability of a given physical system, to
act as a quantum computer using only its natural available
interactions. Therefore the quantum computers are rather a
collection of interacting cells (e.g. quantum dots, nuclear spins,
Josephson junctions etc). These cells are controlled by external
classical switches and they evolve in time by modifying the
switches. The quantum algorithms are translated into  time
manipulations of the external classical switches which control the
system. This kind of quantum computer does not have  connections,
which is the difficult part of a physical implementation. Any
device operating by external classical switches has an internal
range of capabilities, i.e. it can manipulate the quantum
information encoded in a subspace of the full system of Hilbert
space. This capability is called encoded universality of the
system \cite{DiVincpre01}. The notion of the encoded universality
is identical to the notion of the controllability
\cite{RaViMoKo00} of the considered quantum device. The
controllability on Lie groups from a mathematical point of view
was studied in \cite{JurSus75,Rama95,Rama01,KhaGla01,FuSchSo01}.
The controllability of atomic and molecular systems was studied by
several authors see review article \cite{RaViMoKo00} and the
special issue of the \emph{Chemical Physics} vol. 267, devoted to
this problem.  In the case of laser systems the question of
controllability was studied by several authors and recently in
\cite{FuSchSo01,SchFuSo01}. In the present paper we investigate
the conditions in order to obtain the full system of Hilbert space
by a finite number of choices of the values of the classical
switches. As a working example we use the Josephson junction
devices in their simplest form \cite{NakaPRL97, AveSSC98, AveNa99,
ShnSchPRL97} , but this study can be extended in the case of
quantum dots or NMR devices.

In this paper  the intrinsic interaction of a system operating as
a universal computer is employed instead of forcing the system to
enact a predetermined set of universal gates \cite{DiVincpre01}.
 As a basic
building block we use a system of two identical Josephson
junctions coupled by a mutual inductor.  The values of the
classical control parameters (the charging energy $E_{c}$ and the
inductor energy $E_L$) are chosen in such a way that   four basic
Hamiltonians, $H_{i}$,($i=1,\ldots,4$) are created by switching on
and off the bias voltages and the inductor, where the tunneling
amplitude $E_J$ is assumed to be fixed. Our procedure allows the
construction of any one-qubit and two-qubit gate, through a finite
number of steps evolving in time according to the four basic
Hamiltonians. Using the two Josephson junctions network  a
construction scheme of four steps for the simulation is presented
in the case of  an arbitrary one-qubit gate, and of fifteen steps,
for the simulation of an arbitrary two-qubit gate. The main idea
of this paper is the use of a small number of Hamiltonian states
in order to obtain the gates.

Our results can be formulated in terms of a \emph{quantum machine
language}. For example, each two-qubit gate $U$ is represented by
a \emph{command} of the language containing 15 \emph{letters} i.e.
the fifteen steps mentioned above. Each letter consists of a
binary part (the states of the on-off switches), which determines
the basic Hamiltonian used, and a numerical part corresponding to
the time interval.

Our proposal can be  generalized for $N$-qubit gates ($N>2$)
belonging to the $SU(2^{N})$. In that case  $N+2$ basic
Hamiltonians are needed to implement such a gate. The
generalization of these ideas is under investigation.

The paper is organized as follows: In section \ref{sec:One_qubit}
we present for clarity reasons the formalism of Josephson
junctions one-qubit devices. In section \ref{sec:Two_qubit} we
apply the formalism to two qubit gates and simulate two-qubit
gates and the possible $N$-qubit generalizations are discussed. In
section \ref{sec:summary} the results are summarized. Finally in
the Appendix \ref{sec:appendix}, the essential mathematical
feedback is provided.

\section{One-qubit devices}\label{sec:One_qubit}

\begin{figure}[ht]
\centering
\includegraphics[height=1.0in]{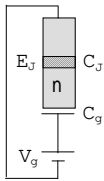}
\caption{  One qubit device}\label{fig:One_qbit}
\end{figure}

The simplest Josephson junction one qubit device is shown in Fig
\ref{fig:One_qbit}. In this section we give a summary of the
considered device. The detailed description and the complete list
of references can be found in the detailed review paper
\cite[section II]{MahkRMP00}. The device consists of a small
superconducting island ("box"), with $n$ excess Cooper pair
charges connected by a tunnel junction with capacitance $C_{J}$
and Josephson coupling energy $E_{J}$ to a superconducting
electrode. A control gate voltage $V_{g}$ (ideal voltage source)
is coupled to the system via a gate capacitor $C_{g}$.

The chosen material is such that the superconducting energy gap is
the largest energy in the problem, larger even than the
single-electron charging energy. In this case quasi-particle
tunneling is suppressed at low temperatures, and a situation can
be reached where no quasi-particle excitation is found on the
island. Under special condition described in ref \cite{MahkRMP00}
only Cooper pairs tunnel coherently in the superconducting
junction.

The voltage $V_g$ is constrained in a range interval where the
number of Cooper pairs takes the values $0$ and $1$, while all
other coherent charge states, having much higher energy, can be
ignored. These charge states correspond to the spin basis
states:\\
\begin{tabular}{l}
  $|\uparrow>$ corresponding to $0$ Cooper-pair charges \\
  on
the island, and \\
  $|\downarrow>$ corresponding to $1$ Cooper-pair charges.
\end{tabular}
\\
In this case the superconducting charge box reduces to a two-state
quantum system, \emph{qubit}, with Hamiltonian (in spin 1/2
notation):
\begin{equation}\label{eq:H_basic}
\begin{array}{c}
 H=\frac{1}{2}E_{c}\,\sigma_{z}-\frac{1}{2}E_{J}\,\sigma_{x},
 \quad \sigma_z |\uparrow>= |\uparrow> \quad \mbox{and} \\ \quad
 \sigma_z |\downarrow>=- |\downarrow>.
\end{array}
\end{equation}
In this Hamiltonian there are two parameters the \emph{bias
energy} $E_c$ and the \emph{tunneling amplitude} $E_J$. The bias
energy $E_c$ is controlled by the gate voltage $V_g$ of Fig
\ref{fig:One_qbit}, while the tunneling amplitude $E_J$ here is
assumed to be constant i.e. it is a constant system parameter. The
tunneling amplitude can be controlled in the case of the tunable
effective Josephson junction, where the single Josephson junction
is replaced by a flux-threaded SQUID \cite{MahkRMP00}, but this
device is more complicated than the one considered in this paper.

The Hamiltonian is written as:
\begin{equation}\label{eq:H_eta}
 H=\frac{1}{2}\Delta E(\eta) (\cos\eta\,\sigma_{z}
 -\sin\eta\,\sigma_{x})
\end{equation}
where $\eta$ is the mixing angle
\[ \eta\equiv \tan^{-1}\frac{E_{J}}{E_{c}}  \]
The energy eigenvalues are
\[ E_\pm= \pm \frac{\Delta E(\eta)}{2}  \]
 and the splitting between the
eigenstates is:
\[ \Delta E(\eta)=\sqrt{E_{J}^2+E_{c}^2}  \]

The eigenstates provided by the Hamiltonian (\ref{eq:H_basic}),
are denoted in the following as $|+>$ and $|->$:
\begin{equation}\label{eq:idle basis}
 \begin{array}{rl}
 |+> &=
 \cos{\frac{\eta}{2}}\,|\downarrow>+ \sin{\frac{\eta}{2}}\,|\uparrow> \\
 |-> &=\sin{\frac{\eta}{2}}\,|\downarrow> -\cos{\frac{\eta}{2}}\,|\uparrow>
 \end{array}
\end{equation}

To avoid confusion we introduce a second set of Pauli matrices
$\vec{\rho}= (\rho_{x},\rho_{y},\rho_{z})$, which operate in the
basis $|+>$, $|->$, while reserving the $\vec{\sigma}$ operators
for the basis of $|\uparrow>$ and $|\downarrow>$:
\[
\begin{array}{c}
 \rho_z=|+><+|\, -\,  |-><-|,\\ \rho_x= |+><-|\,+\,|-><+|, \\
 \rho_y= i |-><+|\, -\, i|+><-|
\end{array}
\]

In  the proposed model we assume that the device of Fig
\ref{fig:One_qbit} has a switch taking two values $1$ and $0$,
corresponding to the switch states \emph{ON} and \emph{OFF}. This
switch controls the gate voltage $V_g$, which takes only two
values either $V_{\rm  id}$ or $V_{\rm  deg}$, where the first one
corresponds to the \emph{ idle Hamiltonian}, while the second one
corresponds  to the \emph{degenerate Hamiltonian}.

The \emph{idle}  point can be achieved for a characteristic value
of the control gate voltage $V_g=V_{\rm  id}$, corresponding to a
special value of the bias energy and to the phase parameter
$\eta=\eta_{\rm  id}$. At this point  the energy splitting $\Delta
E(\eta)$ achieves its maximum value, which is denoted by $ \Delta
E $.
 For simplicity reasons we reserve  the symbol $E_c$ for the
bias energy corresponding to the idle point and   by definition,
the Hamiltonian at the \emph{idle point} then becomes:
\begin{equation}\label{eq:H_idle}
 H_{\rm  id}=\frac{E_c}{2} \sigma_z -\frac{E_J}{2}
 \sigma_x=\frac{1}{2}\Delta E\,\rho_{z}
\end{equation}

At the \emph{degeneracy} point $\eta=\frac{\pi}{2}$ the energy
splitting reduces to $E_{J}$, which is the minimal energy
splitting. This point is characteristic for the material of  the
Josephson junction and corresponds to a special characteristic
choice of the control gate voltage $V_g=V_{\rm  deg}$.
\begin{equation}\label{eq:H_deg}
 H_{\rm  deg}= - \frac{E_{J}}{2} \sigma_x=-\frac{E_{J}}{2} \left(
 \sin \eta_{\rm  id} \rho_z - \cos \eta_{\rm  id} \rho_x \right)
\end{equation}

The system is switched in the state \emph{OFF} (or $0$)
corresponding to the \emph{degenerate} Hamiltonian
(\ref{eq:H_deg}) during a time interval $t_1$, then the system is
switched to the state \emph{ON}  (or $1$) i.e. the \emph{idle}
Hamiltonian (\ref{eq:H_idle}) during a time interval $t_2$ and it
comes back to the initial \emph{degenerate} Hamiltonian during the
time $t_3$. The general form of the evolution operator is:
\begin{equation}\label{eq:One_qbit_Evol}
U= {\rm  e}^{-i t_3 H_{\rm  deg} }  \, {\rm  e}^{-i t_2 H_{\rm id}
} \, {\rm  e}^{-i t_1 H_{\rm  deg} }
\end{equation}

The operators $H_{\rm  id}$ and $H_{\rm  deg}$  and their
commutator
\[
  \left[H_{\rm  id},H_{\rm  deg} \right]= i \frac{E_c E_J}{2} \sigma_y
\]
form a (non-orthogonal) basis of the algebra $su(2)$. Therefore
the pair $H_{\rm  id},\, H_{\rm  deg}$ generates $su(2)$ by taking
these elements and all their possible commutators and their linear
combinations. That means that the combination of three terms as in
equation (\ref{eq:One_qbit_Evol})  for all the triples
$\{t_1,t_2,t_3\}$ cover all the matrices belonging in $SU(2)$.
Thus we conclude that every 2$\times$2 matrix $U$ in $SU(2)$ can
be achieved by a device as in Fig \ref{fig:One_qbit}, with
manipulation of the binary switch permitting to the Hamiltonian
two possible states i.e. the \emph{idle} one and the
\emph{degenerate} one. The above described manipulations can be
codified by a rudimentary \emph{Quantum Machine Language} (QML)
for the one qubit device. In that elementary language the gate $U$
corresponds to a \emph{command} of the language, each command is
constituted by (three) letters, each of them having the form of a
pair
\[
  \{e,t\}, \quad e=0\, (OFF),\mbox{ or } 1\, (ON)
  ,\mbox{ and }  0\le t < \infty
\]
i.e. the command corresponding to equation
(\ref{eq:One_qbit_Evol}) is analyzed in the following (at most
three) letters:
\[
  \begin{array}{|c|}
  \hline U \\
  \hline\hline
  \{0,t_1\}\\
  \hline \{1,t_2\}\\
  \hline \{0,t_3\}\\
  \hline
  \end{array}
\]
The one qubit gates are $2\times 2$ unitary matrices belonging to
the group U(2). Each element in the group $U(2)$  can be projected
up to one multiplication constant to an element of group $SU(2)$.
Evidently the elements generated by the evolution operator
(\ref{eq:One_qbit_Evol}) belong to $SU(2)$. Throughout this paper
we shall use projections of $U(2^N)$ matrices in $SU(2^N)$, using
the symbol ``$\mapsto$'' to denote this projection. Let us
consider the fundamental one qubit gates or commands NOT,
$\sqrt{\rm  NOT}$, Hadamard, Phase Shift and their $SU(2)$
projections:
\[
  {\rm  NOT}=\left(\begin{array}{cc}0 & 1 \\ 1
  &0\end{array}\right)
  \mapsto i \sigma_x = {\rm
  e}^{- i \frac{\pi}{E_{J}}\,H_{\rm  deg} } \in SU(2),
\]
\[
\begin{array}{rl}
  \sqrt{\rm  NOT}=&\frac{1}{\sqrt{2}} \left(\begin{array}{cc}{\rm  e}^{-i \frac{\pi}{4}} & {\rm
  e}^{i \frac{\pi}{4}} \\   {\rm  e}^{i \frac{\pi}{4}} & {\rm  e}^{-i
  \frac{\pi}{4}} \end{array}\right) \mapsto \\
  &\mapsto \frac{1}{\sqrt{2}} \left(\,\mathbb{I}+i
  \sigma_{x} \right)={\rm  e}^{- i \left( \frac{\pi}{2 E_J}\right)\,
  H_{\rm  deg}}
\end{array}
\]
\[
\begin{array}{rl}
  {\rm  Had}=&
  \frac{1}{\sqrt{2}}\left(\begin{array}{cc}1 & 1 \\ 1  &-1\end{array} \right)\mapsto \frac{i}{\sqrt{2}} \left(
  \sigma_x + \sigma_z \right)= \\
  =&{\rm  e}^{-i t_3^{\rm h} H_{\rm  deg}}\,
  {\rm  e}^{-i t_2^{\rm h} H_{\rm  id} }\,{\rm e}^{-i t_1^{\rm h} H_{\rm
  deg}}
\end{array}
\]
where
\[
  t_1^{\rm h}=t_3^{\rm h}= \frac{2 \left( \cos^{-1}\left[ \sqrt{\frac{E_J+E_c}{2
  E_c}}\; \right]+\frac{\pi}{2} \right)}{E_J},
\]
\[
   t_2^{\rm h}=\frac{2 \left( \sin^{-1}
  \left[ {\frac{\Delta E }{\sqrt{2} E_c }}\right]+\pi\right)}{\Delta E}
\]
\[
\begin{array}{rl}
  {\rm PhS}=&\left( \begin{array}{cc} 1 &0 \\ 0 &{\rm  e}^{i \phi}
  \end{array} \right) \mapsto \cos{\frac{\phi}{2}}\,\mathbb{I}-i
  \sin{\frac{\phi}{2}}\,\sigma_{z}=\\
  =&{\rm  e}^{-i t_3^{\rm ph} H_{\rm  deg}}\,
  {\rm  e}^{-i t_2^{\rm ph} H_{\rm  id} } \,{\rm  e}^{-i t_1^{\rm ph} H_{\rm
  deg}}
\end{array}
\]
where
\[
\begin{array}{rl}
  t_1^{\rm ph}=&t_3^{\rm ph}=\\
  =&\frac{2 \left( \cos^{-1}\left[
  \frac{1}{2}\sqrt{2-\frac{\sqrt{2}\,\Delta E\,\sqrt{-1+\frac{2
  E_{c}^{2}}{\Delta E^{2}}+\cos\phi}} {E_{c}\,\cos\frac{\phi}{2}}}
  \; \right]+\frac{\pi}{2} \right)}{E_{J}},
\end{array}
\]
\[
  t_2^{\rm ph}=\frac{2 \left(
  \sin^{-1}
  \left[ \frac{\Delta E \, \sin\frac{\phi}{2}}{E_c }\right] \right)}{\Delta E}
\]

The above formulas imply the following analysis of these commands
in letters (Table \ref{Tab:One}).

\begin{table}[ht]
\[
  \begin{array}{cccc}
  \begin{array}{|c|}
  \hline {\rm  NOT} \\
  \hline\hline
  \{0,\pi/E_J\} \\
  \hline -- \\
  \hline -- \\
  \hline
  \end{array}
&
  \begin{array}{|c|}
  \hline \sqrt{{\rm  NOT}} \\
  \hline\hline
  \{0,\pi/2 E_J\} \\
  \hline -- \\
  \hline -- \\
  \hline
  \end{array}
&
  \begin{array}{|c|}
  \hline {\rm  Had} \\
  \hline\hline
  \{0,t_1^{\rm  h}\} \\
  \hline \{1,t_2^{\rm  h}\} \\
  \hline \{0,t_3^{\rm  h}\} \\
  \hline
  \end{array}
&
  \begin{array}{|c|}
  \hline {\rm  PhS} \\
  \hline\hline
  \{0,t_1^{\rm ph}\} \\
  \hline \{1,t_2^{\rm ph}\} \\
  \hline \{0,t_3^{\rm ph}\} \\
  \hline
  \end{array}
  \end{array}
\]
\caption{Letter Analysis of One-qubit gates} \label{Tab:One}
\end{table}
\normalsize

The above analysis of a quantum gate in letters  is rather trivial
in the one qubit case and it was presented for clarity reasons,
but the similar  construction is far from evident and quite
complicated in the $N$-qubit case.

\section{Two-qubit devices}\label{sec:Two_qubit}

\begin{figure}[ht]
\begin{center}
\includegraphics[width=1.5in]{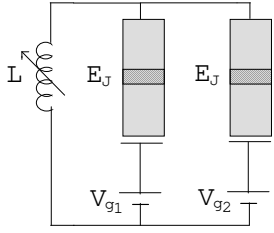}
\caption{   Two qubit device} \label{fig:two_qubit}
\end{center}
\end{figure}

In order to perform one and two qubit quantum gate manipulations
in the same device, we need to couple pairs of qubits together and
to control the interaction between them. For this purpose
identical Josephson junctions are coupled by one mutual inductor
$L$ as shown in Fig \ref{fig:two_qubit}. The physics and the
detailed description of the coupled Josephson junctions are
discussed and reviewed in \cite{MahkRMP00}.
 For $L=0$ the system reduces to
a series of uncoupled, single qubits, while for $L
\rightarrow\infty$ they are coupled strongly. The ideal system
would be one where the coupling between different qubits could be
switched in the state \emph{ON }  (or state $'1'$) by applying an
induction via a constant value inductor $L$ and in the state
\emph{OFF}   (or state $'0'$)  corresponding to $L=0$ and leaving
the qubits uncoupled in the \emph{idle} state.
%%%%%%%%%%%%%%%%%%%
%%%%%%%%
The Hamiltonian for a general two-qubit system is written:
\begin{equation}\label{eq:H2_general}
 \begin{array}{rl}
 H =& H_1+H_2+H_{\rm  int} =\\
 =& {\frac{1}{2}}
 E_{c_1}\,\sigma_z^{(1)}-{\frac{1}{2}}E_{J_1}\,\sigma_x^{(1)}+\\
 &+{\frac{1}{2}}E_{c_2}\,\sigma_z^{(2)}-{\frac{1}{2}}E_{J_2}\,\sigma_x^{(2)}-
 {\frac{1}{2}}E_L\,\sigma_y^{(1)}\,\sigma_y^{(2)}
 \end{array}
\end{equation}
For an explanation of the formalism used in this section see
Appendix \ref{sec:appendix}. In the case of two identical
junctions we have $E_{J_1}=E_{J_2}=E_J$, since the tunneling
amplitude of the junction is a system parameter, depending on the
material. Under these conditions the two coupled Josephson
junctions Hamiltonian will be controlled by the following control
parameters :\,$E_{c_1}, \, E_{c_2}, \,E_L$, which will be called
\emph{switches}. The first two parameters are controlled by the
gate voltages $V_{g_1}, \, V_{g_2}$, while the last parameter is
related to the inductor switch $L$. In the proposed model each of
the parameters $E_{c_1}, \, E_{c_2}$ can have two values $0$ or
$E_c$. The first is the state $'0'$ (or $OFF$) corresponding to
the \emph{degenerate} one qubit state, while the other one is
equal to $E_c$ (or $ON$ or $'1'$ state) corresponding to the one
qubit \emph{idle} state. Also the parameter $E_L$ takes two
values. The one is $E_{L}=0$ ($OFF$ or $'0'$ state) corresponding
to a an uncoupled two qubit state and the other one has a fixed
value ($ON$ or $'1'$ state). For the sake of simplicity we use the
symbol $E_L$ for this induction amplitude. Using this combination
of parameter values or binary switches values, we can obtain the
following four fundamental states of the Hamiltonian
(\ref{eq:H2_general}):
\begin{itemize}
\item[ $H_{1}$:] where both of the junctions are in the idle state $( E_{c_1}=E_{c_2}=E_{c} )$,
 while they are uncoupled $( E_L=0 )$.
\begin{equation}\label{eq:H2/H1=HPP0}
 H_{1} = {\frac{1}{2}} E_{c}\,( \sigma_z^{(1)}+\sigma_z^{(2)})
 -{\frac{1}{2}} E_J\,( \sigma_x^{(1)}+\sigma_x^{(2)} )
\end{equation}
 This Hamiltonian corresponds to the switches choice $(E_{c_1},\,
 E_{c_2}, \, E_L) \rightarrow (1,1,0)$.
\item[ $H_{2}$:] where both of the junctions are in
 the degenerate state $( E_{c_1}=E_{c_2}=0 )$, while the two qubits
 are coupled.
\begin{equation}\label{eq:H2/H2=H00L}
 H_{2} = -{\frac{1}{2}} E_J\,( \sigma_x^{(1)}+\sigma_x^{(2)} )-{\frac{1}{2}}
 E_L\,\sigma_y^{(1)}\sigma_y^{(2)}
\end{equation}
 corresponding to the switch choice $(0,0,1)$.
\item[ $H_{3}$:] where the first junction is in degeneracy $(E_{c_1}=0)$,
 the second is in the idle state $( E_{c_2}=E_{c} )$ and they are
 uncoupled $( E_L=0 )$.
\begin{equation}\label{eq:H2/H3=H0P0}
 H_{3} = {\frac{1}{2}} E_{c}\,\sigma_z^{(2)}-{\frac{1}{2}} E_J\,(
 \sigma_x^{(1)}+\sigma_x^{(2)})
\end{equation}
 corresponding to the switch choice $(0,1,0)$.
\item[ $H_{4}$:] where the first junction is in the idle state
 $( E_{c_1}=E_{c})$,
 the second is in the degeneracy $( E_{c_2}=0)$ and they are
 uncoupled $( E_L=0 )$.
\begin{equation}
 H_{4} = {\frac{1}{2}} E_{c}\,\sigma_z^{(1)}-{\frac{1}{2}} E_J\,(
 \sigma_x^{(1)}+\sigma_x^{(2)}) \label{eq:H2/H3=H0P0a}
\end{equation}
 corresponding to the switch choice $(1,0,0)$.
\end{itemize}
These four Hamiltonian forms  are linearly independent and they
can generate the su$(4)$ algebra by repeated commutations and
linear combinations. For a detailed discussion see the discussion
in Appendix \ref{sec:appendix}, equation (\ref{eq:SigmaToH}). From
Proposition \ref{prop:SU4}, any elementary two-qubit gate, which
is  represented by a unitary $4\times 4$ matrix $U\in$SU$(4)$, can
be constructed as follows:
\begin{equation}\label{eq:impl_uni}
\begin{array}{rl}
 U=&{\rm  e}^{-i H_3 t_{15}} {\rm  e}^{-i H_2 t_{14}} \cdots
   {\rm  e}^{-i H_2 t_6} \cdot  \\
   &\cdot {\rm  e}^{ -i H_1 t_5} {\rm  e}^{-i H_4 t_4}
   {\rm  e}^{  -i H_3 t_3} {\rm  e}^{-i H_2 t_2} {\rm  e}^{ -i H_1 t_1}
\end{array}
\end{equation}
From a physical point of view any two qubit quantum gate can be
obtained by 15 time steps. At the k-th step the device is put at
one of the Hamiltonian states $H_1,\ H_2, \,H_3$ or $H_4$ by
appropriate manipulations of the  switches  during a time interval
$t_k,\, k=1,2,\ldots,15$. Using this fact a \emph{Quantum Machine
Language} (QML) can be defined as in section \ref{sec:One_qubit}.
Each gate, corresponding to the $4\times 4$ matrix  $U$, is
associated to a \emph{command} of the QML. Each command consists
of at most $15$ \emph{letters} or steps, each of them being a
collection of $4$ numbers $\{e_1,\,e_2,\, \ell,\, t\}$ of the
following form:
\[
  \begin{array}{c}
  \mbox{\emph{letter}}\,\Leftrightarrow\, \{e_1,\,e_2,\, \ell,\,t\}
  \qquad 0\le t<\infty \\
  \begin{array}{rll}
  \hline
  e_1= 0 &\mbox{ if } V_{g_1}=V_{\rm  deg} &\; \Rightarrow\; E_{c_1}=0 \\
  = 1 &\mbox{ if } V_{g_1}=V_{\rm  id} &\; \Rightarrow\; E_{c_1}=E_c \\
  \hline
  e_2= 0 &\mbox{ if } V_{g_2}=V_{\rm  deg} &\; \Rightarrow\; E_{c_2}=0 \\
  = 1 &\mbox{ if } V_{g_2}=V_{\rm  id} &\; \Rightarrow\; E_{c_2}=E_c \\
  \hline
  \ell= 0 &\mbox{ if } L=0 &\; \Rightarrow\; E_{L}=0 \\
   = 1 &\mbox{ if }  L\ne 0 &\; \Rightarrow\; E_{L}\ne 0 \\
  \hline
  \end{array}
  \end{array}
\]
The gate $U$ given by equation (\ref{eq:impl_uni}) corresponds to
one \emph{command}, which contains $15$ letters and is presented
in Table \ref{Tab:U}. Each \emph{letter} is the codified command
which indicates the state of the binary switches and the time
interval.
\begin{table}[ht]
\[
  \begin{array}{|c|}
  \hline
  U \\
  \hline
  \hline \{1,1,0,t_1\} \\
  \hline \{0,0,1,t_2\} \\
  \hline \{0,1,0,t_3\} \\
  \hline \{1,0,0,t_4\} \\
  \hline \{1,1,0,t_5\} \\
  \hline \{0,0,1,t_6\} \\
  \hline \{0,1,0,t_7\} \\
  \hline \{1,0,0,t_8\} \\
  \hline \{1,1,0,t_9\} \\
  \hline \{0,0,1,t_{10}\} \\
  \hline \{0,1,0,t_{11}\} \\
  \hline \{1,0,0,t_{12}\} \\
  \hline \{1,1,0,t_{13}\} \\
  \hline \{0,0,1,t_{14}\} \\
  \hline \{0,1,0,t_{15}\} \\
  \hline
  \end{array}
\]
\caption{Letter analysis of a command (gate) $U$} \label{Tab:U}
\end{table}

We should notice that the succession of switch states follows a
cyclic pattern. This regularity might facilitate the manipulation
of the coupled junction device.

Let us now give some numerical simulations of the proposed model
for some fundamental quantum gates essential for the quantum
computation (we use these gates multiplied with a proper constant
because the corresponding matrices should be elements of the
$SU(4)$ group) . These gates are :

\begin{enumerate}
\item The CNOT gate. Probably the most important gate in quantum
computation:
\[
 \begin{array}{cl}
  {\rm  CNOT} &= \left(\begin{array}{cccc}1 &0 &0 &0 \\ 0 &1 &0 &0 \\ 0 &0 &0 &1
  \\
  0 &0 &1 &0 \end{array} \right)\mapsto \\ \\ &\mapsto \frac{{\rm
  e}^{i\frac{\pi}{4}}}{2}
  \,(-\sigma_{z}^{(1)}\sigma_{x}^{(2)}+
  \sigma_{z}^{(1)}+\sigma_{x}^{(2)}+\mathbb{I}\otimes\mathbb{I})
  \in SU(4)
  \end{array}
\]
\item The SWAP gate, which interchanges the input qubits:
\[
  \begin{array}{cl}
  {\rm  SWAP} &=
  \left(\begin{array}{cccc}1 &0 &0 &0 \\   0 &0
  &1 &0 \\   0 &1 &0 &0 \\   0 &0 &0 &1\end{array}\right) \mapsto \\ \\
  &\mapsto \frac{{\rm  e}^{i\frac{\pi}{4}}}{2}\,(\sigma_{x}^{(1)}\sigma_{x}^{(2)}+
  \sigma_{y}^{(1)}\sigma_{y}^{(2)}+\sigma_{z}^{(1)}\sigma_{z}^{(2)}+
  \mathbb{I}\otimes\mathbb{I})
  \end{array}
\]
\item The ${\rm QFT}_4$ gate, the gate of the Quantum Fourier
Transform (the quantum version of the Discrete Fourier Transform),
for 2 qubits. A very useful gate for the implementation of several
quantum algorithms (e.g. Shor's algorithm):
\[
  \begin{array}{rl}
  {\rm  QFT}_4 =&\frac{1}{2} \left(\begin{array}{cccc}1 &1 &1 &1 \\   1 &i &-1 &-i   \\
  1 &-1 &1 &-1 \\   1 &-i &-1 &i  \end{array}\right) \mapsto \\ \\
  &\mapsto \frac{{\rm  e}^{i\frac{3\pi}{8}}}{2\sqrt{2}}(\sigma_{x}^{(1)}\sigma_{z}^{(2)}
  +\mathbb{I}\otimes\mathbb{I})+\frac{{\rm  e}^{-i\frac{\pi}{8}}}{2\sqrt{2}}
  (\sigma_{x}^{(1)}+\sigma_{z}^{(2)})+\\
  &+\frac{{\rm  e}^{i\frac{\pi}{8}}}{2}
  (\sigma_{y}^{(1)}\sigma_{y}^{(2)}+\sigma_{z}^{(1)}\sigma_{x}^{(2)})
  \end{array}
\]
\item A conditional Phase Shift. This gate provides a conditional phase shift
${\rm  e}^{i \phi}$ on the second qubit (adds a phase to the
second qubit):
\[
  \begin{array}{rl}
  {\rm  PhShift} &=
\left(
  \begin{array}{cccc}1 &0 &0 &0 \\   0 &1 &0 &0 \\   0 &0 &1 &0 \\
  0 &0 &0 &{\rm  e}^{i \phi}\end{array} \right) \mapsto \\ \\
  &\mapsto \frac{i \sin{\frac{\phi}{4}}}{2}\,(\sigma_{z}^{(1)}\sigma_{z}^{(2)}
  -\sigma_{z}^{(1)}-\sigma_{z}^{(2)})+\\
  &+\frac{1}{4}\,(3\,{\rm
  e}^{-i\frac{\phi}{4}}+ {\rm  e}^{i\frac{\phi}{4}})\,\mathbb{I}\otimes\mathbb{I}
  \end{array}
\]
 In our analysis we consider this phase to be $\phi
 =\frac{\pi}{2}$.
\end{enumerate}

In the simulations of this paper, the energies are assumed to take
the following values:
\[
E_c= 2.5\, K=3.45\, 10^{-23}\, J, \]
\[ E_J=0.1\, K=0.138\,
10^{-23}\, J, \]
\[E_L=0.1\, K=0.138\, 10^{-23}\, J  \]
i.e. the time scale is of the order of $10^{-11}$ sec . The
numerical value corresponding to the idle state is chosen to
conform to the available experimental data \cite{NakaJLTP00}, and
to keep in the range of different experimental propositions
\cite{ShnSchPRL97, MakhNa99, Mahkpre99, MahkRMP00, NakaNa99}. Each
gate or command $U_{\rm  gate}$ is approximated by an evolution
operator $U(t_1,t_2,\ldots,t_{15})$ of the form
(\ref{eq:impl_uni}). This simulation is equivalent to the analysis
of the command $U_{\rm  gate}$ to letters in conformity with Table
\ref{Tab:U}. The efficiency of our simulation is defined by a
\emph{test function}, $f_{test}$. It is a function of 15 time
variables:
\begin{equation}\label{eq:f_test}
\begin{array}{l}
 f_{test}(t_{1}, t_{2}, \ldots, t_{15})=\\
 =\sum\limits_{i,j=1}^4
 |(U_{\rm  gate})_{ij}-(U(t_{1}, t_{2}, \ldots,
 t_{15}))_{ij}|^2=\\
 =
 || U_{\rm  gate} - U ||^2
\end{array}
\end{equation}
Actually, $f_{test}$ is the norm deviation of our simulation. The
optimum, is obviously the nullification of this norm, $f_{test} =
0$. In fact we apply a minimization procedure and we calculate the
time values, which minimize $f_{test}$. The numerical results are
shown in Table\,\ref{tab:Jj2}.

\begin{table*}[ht]
\[
  \begin{array}{cccc}
  \begin{array}{|c|}
  \hline
  {\rm  CNOT} \\
  \hline
  \hline \{1,1,0,102.7757\} \\
  \hline \{0,0,1,158.8193\} \\
  \hline \{0,1,0,909.9617\} \\
  \hline \{1,0,0,130.0504\} \\
  \hline \{1,1,0,300.7220\} \\
  \hline \{0,0,1,143.9878\} \\
  \hline \{0,1,0,101.0584\} \\
  \hline \{1,0,0,900.5691\} \\
  \hline \{1,1,0,151.5296\} \\
  \hline \{0,0,1,083.4085\} \\
  \hline \{0,1,0,161.0839\} \\
  \hline \{1,0,0,901.9591\} \\
  \hline \{1,1,0,699.3097\} \\
  \hline \{0,0,1,191.4272\} \\
  \hline \{0,1,0,101.2086\} \\
  \hline \hline f_{test}=2.9\times 10^{-6} \\
  \hline
  \end{array}
&
  \begin{array}{|c|}
  \hline
  {\rm  SWAP} \\
  \hline
  \hline \{1,1,0,700.5872\} \\
  \hline \{0,0,1,205.2390\} \\
  \hline \{0,1,0,139.7456\} \\
  \hline \{1,0,0,199.3881\} \\
  \hline \{1,1,0,130.4966\} \\
  \hline \{0,0,1,115.8947\} \\
  \hline \{0,1,0,110.9584\} \\
  \hline \{1,0,0,200.5504\} \\
  \hline \{1,1,0,120.2794\} \\
  \hline \{0,0,1,784.0008\} \\
  \hline \{0,1,0,798.0702\} \\
  \hline \{1,0,0,129.1358\} \\
  \hline \{1,1,0,501.6780\} \\
  \hline \{0,0,1,130.0444\} \\
  \hline \{0,1,0,160.2219\} \\
  \hline \hline f_{test}=9.5\times 10^{-8} \\
  \hline
  \end{array}
&
  \begin{array}{|c|}
  \hline
  {\rm  QFT}_{4} \\
  \hline
  \hline \{1,1,0,710.2581\} \\
  \hline \{0,0,1,084.8635\} \\
  \hline \{0,1,0,159.4397\} \\
  \hline \{1,0,0,142.5689\} \\
  \hline \{1,1,0,133.2760\} \\
  \hline \{0,0,1,653.3505\} \\
  \hline \{0,1,0,133.4825\} \\
  \hline \{1,0,0,924.2883\} \\
  \hline \{1,1,0,173.8055\} \\
  \hline \{0,0,1,633.4525\} \\
  \hline \{0,1,0,701.2262\} \\
  \hline \{1,0,0,131.3048\} \\
  \hline \{1,1,0,849.6562\} \\
  \hline \{0,0,1,128.8483\} \\
  \hline \{0,1,0,150.0286\} \\
  \hline \hline f_{test}=1.0\times 10^{-7}\\
  \hline
  \end{array}
&
  \begin{array}{|c|}
  \hline
  {\rm  Phase\,Shift} (\phi=\frac{\pi}{2}) \\
  \hline
  \hline \{1,1,0,110.7116\} \\
  \hline \{0,0,1,902.6813\} \\
  \hline \{0,1,0,120.3082\} \\
  \hline \{1,0,0,397.8240\} \\
  \hline \{1,1,0,109.4998\} \\
  \hline \{0,0,1,175.8195\} \\
  \hline \{0,1,0,521.3209\} \\
  \hline \{1,0,0,122.5053\} \\
  \hline \{1,1,0,102.2305\} \\
  \hline \{0,0,1,795.3231\} \\
  \hline \{0,1,0,108.9077\} \\
  \hline \{1,0,0,198.7137\} \\
  \hline \{1,1,0,159.0513\} \\
  \hline \{0,0,1,888.0009\} \\
  \hline \{0,1,0,100.7132\} \\
  \hline \hline f_{test}=1.2\times 10^{-7}\\
  \hline
  \end{array}
  \end{array}
\]
 \caption{   Letter analysis of Two-Qubit Gates }\label{tab:Jj2}
%\begin{ruledtabular}
\end{table*}
\normalsize

In our numerical examples we use an approximation of the time
parameters to the fourth decimal digit. Respectively we calculate
the value of the test function. Taking into consideration three
more decimal digits the test function $f_{test}$ attains values of
the order of $10^{-10}$. It is a matter of intensity of the
numerical algorithms used to find the minimum of the test function
($f_{test}=0$) and convention of the number of decimal digits of
the time parameters to succeed the optimal simulation . Indeed
time parameters can not be determined with absolute precision in
an implementation scheme for quantum computation.

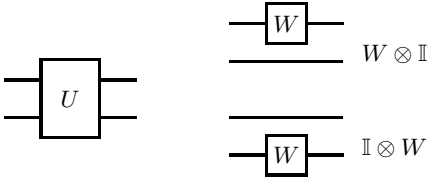
\begin{figure}[ht]
\centering

\begin{picture}(10,5) \thicklines

\put(0.0,1.5){\line(1,0){1.0}}\put(1.0,1.0){\framebox(1.5,2){$U$}}
\put(0.0,2.5){\line(1,0){1.0}}\put(2.5,1.5){\line(1,0){1.0}}
\put(2.5,2.5){\line(1,0){1.0}}

\put(6.0,0.5){\line(1,0){1.0}}\put(7.0,0.0){\framebox(1,1){$W$}}
\put(8.0,0.5){\line(1,0){1.0}}\put(6.0,1.5){\line(1,0){3.0}}
\put(9.5,0.5){$\mathbb{I}\otimes W$}

\put(6.0,3.0){\line(1,0){3.0}}\put(6.0,4.0){\line(1,0){1.0}}
\put(7.0,3.5){\framebox(1,1){$W$}}\put(8.0,4.0){\line(1,0){1.0}}
\put(9.5,3.0){$W\otimes\mathbb{I}$}

\end{picture}
\caption{ Two-qubit and one-qubit gates in two qubit
networks}\label{fig:One_qb_gates}

\end{figure}

The construction of the one-qubit gates with the two-qubit
Josephson device of Fig \ref{fig:two_qubit} is possible. That is
gates of the form $\mathbb{I}\otimes W$ and $W\otimes \mathbb{I}$,
where $W\in SU(2)$, Fig \ref{fig:One_qb_gates}, which are
simulated by the same device.

The construction scheme comes as follows:
\begin{equation}\label{eq:1qbit_gate}
 U={\rm  e}^{-i t_{4} H_{1}}{\rm  e}^{-i t_{3} H_{4}}{\rm  e}^{-i t_{2} H_{1}}
 {\rm  e}^{-i t_{1} H_{4}}
\end{equation}
where $H_{1}$ and $H_{4}$ are special forms of the Hamiltonian
(\ref{eq:H2_general}), rewritten in the idle basis and
$t_{1},\ldots,t_{4}$ the time duration of each step. Obviously:
\[
\begin{array}{rl}
  H_{1}=&\left({\frac{1}{2}} E_{c}\,\sigma_z^{(1)}-{\frac{1}{2}}
  E_J\,\sigma_x^{(1)}\right)+ \\
  &+\left({\frac{1}{2}}
  E_{c}\,\sigma_z^{(2)}-{\frac{1}{2}} E_J\,\sigma_x^{(2)}\right)
  =\frac{\Delta E}{2}(\rho_{z}^{(1)}+\rho_{z}^{(2)})
\end{array}
\]
\[
\begin{array}{rl}
  H_{4}=& \left({\frac{1}{2}} E_{c}\,\sigma_z^{(1)}-{\frac{1}{2}} E_J\,
  \sigma_x^{(1)} \right) -{\frac{1}{2}} E_J\,
  \sigma_x^{(2)}=\\
  =&\frac{\Delta E}{2}\left( \rho_{z}^{(1)}+\tau^{(2)}
  \right)
\end{array}
\]
where
\[ \tau= -\frac{E_J}{\Delta E} \sigma_x  \]
It can easily be shown that:
\[
\begin{array}{l}
  U={\rm  e}^{-i t_{tot}\frac{\Delta E}{2}\rho_{z}} \otimes \\
  \otimes{\rm  e}^{-i t_{4}\frac{\Delta E}{2}\rho_{z}}
  {\rm  e}^{-i t_{3}\frac{\Delta E}{2}\tau} {\rm  e}^{-i t_{2}\frac{\Delta E}{2}\rho_{z}}
  {\rm  e}^{-i t_{1}\frac{\Delta E}{2}\tau}
\end{array}
\]
Setting the total time
\[ t_{tot}=\sum_{i=1}^{4}t_{i}= \frac{4k\pi}{\Delta E}\quad, k\in\mathbb{N} \]
the previous relation is written as:
\[
  U=\mathbb{I} \otimes {\rm  e}^{-i t_{4}\frac{\Delta E}{2}\rho_{z}}
  {\rm  e}^{-i t_{3}\frac{\Delta E}{2}\tau}{\rm  e}^{-i t_{2}\frac{\Delta E}{2}\rho_{z}}
  {\rm  e}^{-i t_{1}\frac{\Delta E}{2}\tau}
\]
The right hand side of the last relation is a $2\times 2$ SU(2)
matrix depending on $3$ independent time parameters $t_1,\, t_2,\,
t_3$, since the fourth time parameter $t_4$ is specified from the
demand that total time is assumed to be fixed. By an appropriate
choice of these three time parameters any gate $U$ of the above
form can be constructed. So any command of the form
$U=\mathbb{I}\otimes W$, which corresponds to an one qubit gate
can be constructed by at most four steps. Therefore any command
$\mathbb{I}\otimes W$ can be analyzed in at most four letters. We
simulate numerically the proposed model for the following
one-qubit gates, $\mathbb{I}\otimes(\rm NOT),\,
\mathbb{I}\otimes(\rm Had),\, \mathbb{I}\otimes(\sqrt{\rm  NOT})
\,{\rm  and}\,\mathbb{I}\otimes PhS$:

\[
  \mathbb{I}\otimes {\rm  NOT}=
  \left( \begin{array}{cccc} 0 &1 &0 &0 \\ 1 &0 &0 &0 \\ 0 &0 &0 &1 \\
  0 &0 &1 &0 \end{array} \right) \mapsto \mathbb{I}\otimes
  (i\sigma_{x}) \in SU(4)
\]
\[
  \mathbb{I}\otimes {\rm  h}=\frac{1}{\sqrt{2}}\left( \begin{array}{cccc}
  1 &1 &0 &0 \\ 1 &-1 &0 &0 \\ 0 &0 &1 &1 \\ 0 &0 &1 &-1
  \end{array} \right) \mapsto \mathbb{I}\otimes \left( \frac{i}{\sqrt{2}}\,
  (\sigma_{x}+\sigma_{z}) \right)
\]
\[\begin{array}{rl}
  \mathbb{I}\otimes \sqrt{\rm  NOT}=&\frac{1}{\sqrt{2}}\left( \begin{array}{cccc}
  {\rm  e}^{-i \frac{\pi}{4}} &{\rm  e}^{i \frac{\pi}{4}} &0 &0 \\
  {\rm  e}^{i \frac{\pi}{4}} &{\rm  e}^{-i \frac{\pi}{4}} &0 &0 \\
  0 &0 &{\rm  e}^{-i \frac{\pi}{4}} &{\rm  e}^{i \frac{\pi}{4}} \\
  0 &0 &{\rm  e}^{i \frac{\pi}{4}} &{\rm  e}^{-i \frac{\pi}{4}}
  \end{array} \right) \mapsto \\
  &\mapsto \mathbb{I}\otimes \left( \frac{1}{\sqrt{2}}
  (\mathbb{I}+i\sigma_{x}) \right)
\end{array}
\]
\[
  \mathbb{I}\otimes {\rm  PhS}=\left( \begin{array}{cccc}
  1 &0 &0 &0 \\ 0 &{\rm  e}^{i \phi} &0 &0 \\ 0 &0 &1 &0 \\
  0 &0 &0 &{\rm  e}^{i \phi} \end{array} \right) \mapsto \mathbb{I}\otimes
  \left( \cos\frac{\phi}{2}\,\mathbb{I}-i\sin\frac{\phi}{2}\,\sigma_{z}
  \right)
\]

   The case of
the gates of the form $U=W \otimes \mathbb{I}$ can be similarly
treated by using basic Hamiltonians $H_1$ and $H_3$.

\[
\begin{array}{rl}
  H_{1}=&\left({\frac{1}{2}} E_{c}\,\sigma_z^{(1)}-{\frac{1}{2}}
  E_J\,\sigma_x^{(1)}\right)+ \left({\frac{1}{2}}
  E_{c}\,\sigma_z^{(2)}-{\frac{1}{2}} E_J\,\sigma_x^{(2)}\right)\\
  =&\frac{\Delta E}{2}(\rho_{z}^{(1)}+\rho_{z}^{(2)})
\end{array}
\]
\[
\begin{array}{rl}
  H_{3}=& -{\frac{1}{2}} E_J\,\sigma_x^{(1)}
  \left({\frac{1}{2}} E_{c}\,\sigma_z^{(2)}-{\frac{1}{2}} E_J\,\sigma_x^{(2)}
  \right)\\
  =&\frac{\Delta E}{2}\left(\tau^{(1)}+\rho_{z}^{(2)}\right)
\end{array}
\]
where
\[ \tau= -\frac{E_J}{\Delta E} \sigma_x  \]

The equivalent construction scheme is:
\begin{equation}\label{eq:1qbit_gate_b}
 U={\rm  e}^{-i t_{4} H_{1}}{\rm  e}^{-i t_{3} H_{3}}{\rm  e}^{-i t_{2} H_{1}}
 {\rm  e}^{-i t_{1} H_{3}}
\end{equation}
Obviously,
\[
\begin{array}{lr}
  U=&{\rm  e}^{-i t_{4}\frac{\Delta E}{2}\rho_{z}}
  {\rm  e}^{-i t_{3}\frac{\Delta E}{2}\tau}
  {\rm  e}^{-i t_{2}\frac{\Delta E}{2}\rho_{z}}
  {\rm  e}^{-i t_{1}\frac{\Delta E}{2}\tau}
  \otimes\\
  &\otimes
  {\rm  e}^{-i t_{tot}\frac{\Delta E}{2}\rho_{z}}
\end{array}
\]
Setting the total time
\[
  t_{tot}=\sum_{i=1}^{4}t_{i}= \frac{4k\pi}{\Delta E}
  ,\quad k\in\mathbb{N}
\]
the previous relation is written as:
\[
  U={\rm  e}^{-i t_{4}\frac{\Delta E}{2}\rho_{z}}
  {\rm  e}^{-i t_{3}\frac{\Delta E}{2}\tau}
  {\rm  e}^{-i t_{2}\frac{\Delta E}{2}\rho_{z}}
  {\rm  e}^{-i t_{1}\frac{\Delta E}{2}\tau} \otimes \mathbb{I} \]
It is apparent that the numerical results for a simulation of an
one-qubit gate should be the same regardless of its form,
$\mathbb{I}\otimes W$ or $W\otimes\mathbb{I}$. So the numerical
results concerning the time parameters presented in Table
\ref{tab:Jj1} should be the same for the simulation of the
corresponding gates $\rm NOT\otimes\mathbb{I},\, \rm
Had\otimes\mathbb{I},\, \sqrt{\rm  NOT}\otimes\mathbb{I},\, {\rm
PhS}\otimes\mathbb{I}$.

\begin{table*}[ht]
\[
  \begin{array}{cccc}
  \begin{array}{|c|}
  \hline \mathbb{I}\otimes {\rm  NOT}
  \\
  \hline\hline \{1,0,0,133.6621\} \\
  \hline \{1,1,0,104.2929\} \\
  \hline \{1,0,0,102.2461\} \\
  \hline \{1,1,0,111.8267\} \\
  \hline \hline f_{test}=1.5\times 10^{-8} \\
  \hline
  \end{array}
  &
  \begin{array}{|c|}
  \hline \mathbb{I}\otimes {\rm Had}
  \\
  \hline\hline \{1,0,0,113.3151\} \\
  \hline \{1,1,0,111.1253\} \\
  \hline \{1,0,0,109.4076\} \\
  \hline \{1,1,0,118.1799\} \\
  \hline \hline f_{test}=1.5\times 10^{-8} \\
  \hline
  \end{array}
  &
  \begin{array}{|c|}
  \hline \mathbb{I}\otimes \sqrt{\rm  NOT}
  \\
  \hline\hline \{1,0,0,115.7315\} \\
  \hline \{1,1,0,114.3695\} \\
  \hline \{1,0,0,100.0236\} \\
  \hline \{1,1,0,121.9033\} \\
  \hline \hline f_{test}=3.4\times 10^{-9} \\
  \hline
  \end{array}
  &
  \begin{array}{|c|}
  \hline \mathbb{I}\otimes {\rm  PhS}
  \\
  \hline\hline \{1,0,0,114.9978\} \\
  \hline \{1,1,0,109.2394\} \\
  \hline \{1,0,0,115.3838\} \\
  \hline \{1,1,0,112.4068\} \\
  \hline \hline f_{test}=1.5\times 10^{-9} \\
  \hline
  \end{array}
  \\
  \\
  \begin{array}{|c|}
  \hline {\rm  NOT}\otimes\mathbb{I}
  \\
  \hline\hline \{0,1,0,133.6621\} \\
  \hline \{1,1,0,104.2929\} \\
  \hline \{0,1,0,102.2461\} \\
  \hline \{1,1,0,111.8267\} \\
  \hline \hline f_{test}=1.5\times 10^{-8} \\
  \hline
  \end{array}
  &
  \begin{array}{|c|}
  \hline {\rm  h}\otimes\mathbb{I}
  \\
  \hline\hline \{0,1,0,113.3151\} \\
  \hline \{1,1,0,111.1253\} \\
  \hline \{0,1,0,109.4076\} \\
  \hline \{1,1,0,118.1799\} \\
  \hline \hline f_{test}=1.5 \times 10^{-8} \\
  \hline
  \end{array}
  &
  \begin{array}{|c|}
  \hline \sqrt{\rm  NOT}\otimes\mathbb{I}
  \\
  \hline\hline \{0,1,0,115.7315\} \\
  \hline \{1,1,0,114.3695\} \\
  \hline \{0,1,0,100.0236\} \\
  \hline \{1,1,0,121.9033\} \\
  \hline \hline f_{test}=3.4\times 10^{-9} \\
  \hline
  \end{array}
  &
  \begin{array}{|c|}
  \hline {\rm  PhS}\otimes\mathbb{I}
  \\
  \hline\hline \{0,1,0,114.9978\} \\
  \hline \{1,1,0,109.2394\} \\
  \hline \{0,1,0,115.3838\} \\
  \hline \{1,1,0,112.4068\} \\
  \hline \hline f_{test}=1.5\times 10^{-9} \\
  \hline
  \end{array}
  \end{array}
\]
 \caption{   Letter analysis of one-qubit gates in two qubit networks}
 \label{tab:Jj1}
\end{table*}

Here it must be  noticed that usually in order to achieve the
construction of one qubit gates a more complicated technique is
usually proposed. In this method the Josephson junctions are
``neutralized'' by appropriate annihilation of the tunneling
amplitude $E_J$ by using SQUID techniques \cite{MakhNa99,
MahkRMP00}.

\section{Summary}\label{sec:summary}

The traditional approach to quantum computing is the construction
of elementary one-qubit and two-qubit gates (universal set of
quantum gates) which are connected by quantum connections and can
represent any quantum algorithm \cite{BaBePRA95}. A different view
is employed in the present paper, proposed  in
\cite{DiVincpre01,DiVincNa00} under the name of encoded
universality. According to this, we do not force the system to act
as a predetermined set of universal gates connected by quantum
connections, but we exploit its intrinsic ability to act as a
quantum computer employing its natural available interaction.

Thus,  any one-qubit and two-qubit gate can be expressed by two
identical Josephson junctions coupled by a mutual inductor. This
can be realized by a finite number of time steps evolving
according to a restricted collection of basic Hamiltonians. These
Hamiltonians are implemented using the above system of junctions
by choosing suitably the control parameters, by switching on and
off the bias voltages and the mutual inductor. The interaction
times of the steps are calculated numerically.

The values of the switches together with the values of the time
steps may constitute the \emph{quantum machine language}. Each
\emph{command} of the language consists of a series of
\emph{letters} and each letter of a binary part (the values of the
switch characterizing the Hamiltonian) and a numerical part (the
interaction time).

The generalization to $N$-qubit gates is currently under
investigation. In this case we need $N+2$ basic Hamiltonians in
order to represent the corresponding $N$-qubit gate. The structure
of commands is an open problem. Each command can be obtained by
$2^N-1$ letters see Proposition \ref{prop:SU8} in Appendix
\ref{sec:appendix}. The mathematical foundation of this conjecture
will be studied in another specialized paper. However by using the
techniques described in \cite{BaBePRA95}, the number of letters
can be reasonably reduced in the $N$-qubit case. The application
of the same methodology for other devices as quantum dots and NMR
are under investigation.

\begin{appendix}
\section[Appendix]{Minimal generating set of the su$(2^N)$ - algebra}
\label{sec:appendix}
 Let us consider the su$(2)$ algebra in the
adjoint representation. This algebra representation is a three
dimensional vector space with basis the $2\times 2$ Pauli
matrices:
\[
  \sigma_z= \left( \begin{array}{cc}1  &0\\0 & 1 \end{array}
  \right), \quad
  \sigma_x= \left(\begin{array}{cc} 0 &1\\1 & 0 \end{array} \right)
  \quad \mbox{ and }\quad
  \sigma_y= \left(\begin{array}{cc} 0 &i\\-i & 0 \end{array} \right)
\]
therefore
\[ {\rm  su}(2) = {\rm  span}(\sigma_z, \sigma_x, \sigma_z ) \]

This su$(2)$ algebra in the adjoint representation is generated by
the following  $2\times 2$ hermitian matrices:
\[
  \sigma_z= \left( \begin{array}{cc}1  &0\\0 & 1 \end{array}
  \right)\quad \mbox{ and }\quad
  \sigma_x= \left(\begin{array}{cc} 0 &1\\1 & 0 \end{array} \right)
\]
because
\[ \left[ \sigma_z, \sigma_x \right]= 2 i \sigma_y  \]
The adjoint representation of the algebra su$(2^2)$ is the vector
space spanned by the 15 matrices
\begin{equation}\label{eq:bas_su4}
 {\rm  su}(2^2) = {\rm  span}( \sigma_i^{(1)},\, \sigma_i^{(2)}, \,
 \sigma_i^{(1)} \sigma_j^{(2)},\, i,j=1,2,3 )
\end{equation}
where
\[
  \sigma_i^{(1)}=\sigma_i \otimes \mathbb{I}, \quad
  \sigma_i^{(2)}=\mathbb{I} \otimes \sigma_i,\quad
  \sigma_i^{(1)} \sigma_j^{(2)} =\sigma_i \otimes \sigma_j
\]
the adjoint representation of the su$(2^2)$ algebra can be
generated by linear combinations and successive commutations of
the following 4 elements:
\begin{equation}\label{eq:gen_su4}
 \sigma_z^{(1)}, \quad \sigma_z^{(2)},\quad
 \sigma_x^{(1)}+\sigma_x^{(2)} , \mbox{  and }
 \sigma_y^{(1)}\sigma_y^{(2)}
\end{equation}
This is indeed true because  all the elements of the basis
(\ref{eq:bas_su4}) can be generated by repeated commutations of
the elements (\ref{eq:gen_su4}), because for $k=1,2,3$:
\begin{equation}\label{eq:constr_lin}
 \begin{array}{rl}
 \sigma_y^{(k)}=&\frac{i}{2}
 \left[\sigma_x^{(1)}+\sigma_x^{(2)},\sigma_z^{(k)}\right], \\
 \sigma_x^{(k)}=& \frac{i}{2} \left[\sigma_z^{(k)}, \sigma_y^{(k)}
 \right]=\\
 =&
 \frac{1}{4}\left[\sigma_z^{(k)},\left[\sigma_z^{(k)},\sigma_x^{(1)}+\sigma_x^{(2)}
 \right]\right]
 \end{array}
\end{equation}
The elements $\sigma_k^{(1)} \sigma_\ell^{(2)}$ can be generated
by commutating  the generators $\sigma_m^{(i)}$ with
$\sigma_y^{(1)}\sigma_y^{(2)}$. One illustrative example is the
construction of the element $\sigma_x^{(1)}\sigma_z^{(2)}$,  by
using the generators (\ref{eq:gen_su4}):
\begin{equation}\label{eq:constr_qua}
 \begin{array}{l}
 \sigma_x^{(1)}\sigma_z^{(2)}= \frac{1}{4}
 \left[\sigma_x^{(2)},\left[\sigma_z^{(1)},\sigma_y^{(1)}\sigma_y^{(2)}\right]
 \right]=\\
 = \frac{1}{16} \left[
 \left[\sigma_z^{(k)},\left[\sigma_z^{(k)},\sigma_x^{(1)}+\sigma_x^{(2)}
 \right]\right],
 \left[\sigma_z^{(1)},\sigma_y^{(1)}\sigma_y^{(2)}\right]\right]
 \end{array}
\end{equation}
In the case of the adjoint representation of the algebra
su$(2^3)$, we can work following a similar methodology. The
adjoint representation algebra su$(2^3)$ is a vector space spanned
by the following 63 matrices:
\begin{equation}\label{eq:bas_su8}
\begin{array}{rl}
 {\rm  su}(2^3) =& {\rm  span}( \sigma_i^{(1)},\, \sigma_i^{(2)},
 \,\sigma_i^{(3)}, \, \sigma_i^{(1)} \sigma_j^{(2)},\\
 &\sigma_i^{(1)} \sigma_j^{(2)} \sigma_k^{(3)},\, i,j,k=1,2,3 )
\end{array}
\end{equation}
where
\[
  \sigma_i^{(1)}=\sigma_i \otimes \mathbb{I} \otimes \mathbb{I},
  \quad \sigma_i^{(2)}=\mathbb{I} \otimes \sigma_i\otimes \mathbb{I}
  \quad \sigma_i^{(3)}=\mathbb{I} \otimes \mathbb{I}\otimes \sigma_i
\]
The above elements can be generated by repeated commutations of
the following 5 matrices:
\begin{equation}\label{eq:gen_su8}
 \begin{array}{c}
 \sigma_z^{(1)}, \quad \sigma_z^{(2)},\quad \sigma_z^{(3)},\\
 \sigma_x^{(1)}+\sigma_x^{(2)}+\sigma_x^{(3)}   , \mbox{  and } \\
 \sigma_y^{(1)}\sigma_y^{(2)}+
 \sigma_y^{(1)}\sigma_y^{(3)}+\sigma_y^{(2)}\sigma_y^{(3)}
 \end{array}
\end{equation}
The linear terms $\sigma_i^{(k)}$ can be easily generated by
formulas as in equation (\ref{eq:constr_lin}). The quadratic terms
$\sigma_i^{(k)} \sigma_j^{(\ell)}$ are generated by manipulations
slightly more complicated than in the case of equation
(\ref{eq:constr_qua}). Let us take the example of the generation
of the element $\sigma_x^{(1)}\sigma_z^{(2)}$, then we must
perform the following commutation actions:
\begin{equation}\label{eq:constr_qua1}
 \begin{array}{l}
 \sigma_y^{(1)}\sigma_x^{(2)}+\sigma_x^{(2)}\sigma_y^{(3)}=\\
 =
 \frac{i}{2}\left[\sigma_z^{(2)},\sigma_y^{(1)}\sigma_y^{(2)}+
 \sigma_y^{(1)}\sigma_y^{(3)}+\sigma_y^{(2)}\sigma_y^{(3)}\right]\\
 \sigma_x^{(1)}\sigma_x^{(2)}=
 \frac{i}{2}\left[\sigma_z^{(1)},\sigma_y^{(1)}\sigma_x^{(2)}+\sigma_x^{(2)}\sigma_y^{(3)}\right]\\
 \sigma_y^{(2)}= \frac{i}{2}\left[\sigma_z^{(2)},
 \sigma_x^{(1)}+\sigma_x^{(2)}+\sigma_x^{(3)} \right]\\
 \sigma_x^{(1)}\sigma_z^{(2)}=\frac{i}{2} \left[ \sigma_y^{(2)},
 \sigma_x^{(1)}\sigma_x^{(2)} \right]
 \end{array}
\end{equation}
Therefore all the quadratic terms can be generated by the elements
(\ref{eq:gen_su8}). Let us now generate a cubic term of the
algebra as the element
$\sigma_z^{(1)}\sigma_y^{(2)}\sigma_x^{(3)}$. This element is
generated by the commutation elements
$\sigma_z^{(1)}\sigma_z^{(2)}$ and $\sigma_x^{(2)}\sigma_x^{(3)}$,
which are generated previously:
\[
  \sigma_z^{(1)}\sigma_y^{(2)}\sigma_x^{(3)}= \frac{i}{2}
  \left[\sigma_z^{(1)}\sigma_z^{(2)},\sigma_x^{(2)}\sigma_x^{(3)}\right]
\]

By induction we can prove the following proposition:
\begin{prop}\label{prop:su2N}
The adjoint hermitian representation of the algebra su$(2^N)$, i.e
the set of hermitian traceless $2^N \times 2^N$ can generated by
the algebra of Lie-polynomials of the set:
\begin{equation}\label{eq:gen_su2N}
 {\mathcal A}_N = \left\{
 \sigma_z^{(1)},\,\sigma_z^{(2)},\,\ldots,\, \sigma_z^{(N)},\,
 \sum\limits_{k=1}^{N} \sigma_x^{(k)},\, \sum\limits_{i<j}^{N}
 \sigma_y^{(i)}\sigma_y^{(j)} \right\}
\end{equation}
\end{prop}
The set ${\mathcal A}_N$ of the generators has $N+1$ elements, we
should point out that this number is very small than the number
$4^N-1$, which is the dimension of the algebra su$(2^N)$.
Therefore, large Lie algebras can be generated by using a
relatively small number of elements.

Let us now construct the group SU$(2^N)$. For the sake of
simplicity we start the discussion with the SU$(4)$ case, i.e with
the set of unitary $4\times 4$ matrices with determinant equal to
$1$.

Let us consider four linearly independent elements, which are
given by the formulas:
\begin{equation}\label{eq:H_basis}
 \begin{array}{l}
 H_{1} = {\frac{1}{2}} E_{\rm  c}\,( \sigma_z^{(1)}+\sigma_z^{(2)}
 )-{\frac{1}{2}} E_J\,( \sigma_x^{(1)}+\sigma_x^{(2)} )
 \\
 H_{2} = -{\frac{1}{2}} E_J\,( \sigma_x^{(1)}+\sigma_x^{(2)} )
 -{\frac{1}{2}} E_L\,\sigma_y^{(1)}\sigma_y^{(2)}
 \\
 H_{3} = {\frac{1}{2}} E_{\rm  c}\,\sigma_z^{(2)}-{\frac{1}{2}} E_J\,(
 \sigma_x^{(1)}+\sigma_x^{(2)})
 \\
 H_{4} = {\frac{1}{2}} E_{\rm  c}\,\sigma_z^{(1)}-{\frac{1}{2}} E_J\,(
 \sigma_x^{(1)}+\sigma_x^{(2)})
 \end{array}
\end{equation}
Starting from this system we can reconstruct the elements
(\ref{eq:gen_su4}) because:
\begin{equation}\label{eq:SigmaToH}
 \begin{array}{l}
 \sigma_z^{1}= \frac{2 (H_1-H_3) }{E_{\rm  c}}\\
 \sigma_z^{2}= \frac{2 (H_1-H_4) }{E_{\rm  c}}\\
 \sigma_x^{1} +\sigma_x^{2}= \frac{ 2 ( H_1-H_3 - H_4 ) }{E_{J }}\\
 \sigma_y^{1} \sigma_y^{2}= \frac{ 2 ( H_3+ H_4 - H_1 - H_2 ) }{E_L}\\
 \end{array}
\end{equation}
These relations prove that the su$(4)$ algebra can be generated by
combinations and successive commutations of the four elements
$\{H_1,\, H_2,\, H_3,\, H_4 \}$. Starting from this fact we can
construct all the elements of the form:
\begin{equation}\label{eq:U_U4}
\begin{array}{rl}
 U=&{\rm  e}^{-i H_3 t_{15}}{\rm  e}^{-i H_2 t_{14}} \cdots {\rm  e}^{-i H_2 t_6}
   {\rm  e}^{-i H_1 t_5}{\rm  e}^{-i H_4 t_4}\cdot\\
   &\cdot{\rm  e}^{-i H_3 t_3}{\rm  e}^{-i H_2 t_2}
   {\rm  e}^{-i H_1 t_1}
\end{array}
\end{equation}
From the Baker-Campbell-Hausdorff (BCH) formula \cite[Sec.
2.15]{Varadarajan}:
\[
  \begin{array}{c}
  {\rm  e}^{-i A}{\rm  e}^{-i B}={\rm  e}^{-i h(A,B)},\\
  \quad \mbox{where}\quad \\
  \begin{array}{rl}
  h(A,B)=&A+B+\frac{i}{2}[A,B] -\\&-\frac{1}{12}\left(
  [[A,B],B]-[[A,B],A]\right)- \\
  &-\frac{i}{48}\left( [B,[A,[A,B]]]-[A,[B,[B,A]]]\right)+\ldots
  \end{array}
  \end{array}
\]
one can calculate $U$ in (\ref{eq:U_U4}) by successive
applications of the the above BCH formula starting from the left
to the right, i.e.
\[
  U=\exp[{-iu_{15}}] \quad \mbox{where} \quad
  \begin{array}[t]{l}
  u_1= t_1 H_1,\\
  u_{2}=h( t_2 H_2,u_1),\\ u_{3}=h( t_3 H_3,u_2) \\
  \qquad \ldots\ldots\\
  u_{15}=h(t_{15}  H_3,u_{14})
  \end{array}
\]
The elements $u_k$ are complicated combinations of the elements
$H_1,\, H_2,\, H_3,\, H_4 $, bracketed inside commutators, i.e are
Lie-polynomials of the free associative algebra
$\mathbb{C}\{H_1,\, H_2,\, H_3,\, H_4\}$. By definition the
combinations and all the successive commutators of these elements
generate the algebra su$(4)$, which has as a linear basis the 15
elements given by equation (\ref{eq:bas_su4}). Thus the general
form of $u_{15}$ is given by:
\[
  u_{15}= \sum\limits_{i=1}^{3} \left(f^{(1)}_i \sigma_i^{(1)} +
  f^{(2)}_i \sigma_i^{(2)} \right) + \sum\limits_{i,j=1}^{3}
  f^{(1,2)}_{ij} \sigma_i^{(1)}\sigma_j^{(2)}
\]
where the 15 coefficients
\begin{equation}\label{eq:f_functions}
\begin{array}{c}
 f^{(k)}_i=f^{(k)}_i(t_1,t_2,\ldots,t_{15})\\ \mbox{and} \\
 f^{(1,2)}_i=f^{(1,2)}_i(t_1,t_2,\ldots,t_{15})
\end{array}
\end{equation}
are complicated functions of the parameters
$t_1,t_2,\ldots,t_{15}$. Any element of the algebra su$(4)$ is
written as a linear combination of the elements of the basis
(\ref{eq:bas_su4}) and from the known functions
(\ref{eq:f_functions}) we can find the values  of the finite time
series $t_1,\,t_2,\ldots,t_{15}$. Then we have proved the
following Proposition

\begin{prop}\label{prop:SU4}
The group SU$(4)$ is given by the elements of the form:
\[\begin{array}{rl}
  U=&{\rm  e}^{-i H_3 t_{15}}{\rm  e}^{-i H_2 t_{14}} \cdots {\rm  e}^{-i H_2
  t_6}\cdot \\
  &\cdot
  {\rm  e}^{-i H_1 t_5}{\rm  e}^{-i H_4 t_4}{\rm  e}^{-i H_3 t_3}{\rm  e}^{-i H_2 t_2}
  {\rm  e}^{-i H_1 t_1}
\end{array}
\]
where $H_1, \, H_2,\, H_3,\, H_4$ are some special elements of the
algebra su$(4)$. The combinations and the successive commutations
of these elements generate su$(4)$.
\end{prop}
This proposition is a special form of the bang-bang
controllability theory for SU$(4)$ matrices \cite{JurSus75}. The
above decomposition of the SU$(4)$ matrices is a generalization of
the Euler decomposition of the SU$(2)$ matrices. In
\cite{KhaGla01} another decomposition is proposed based on the
Cartan decomposition of SU$(2^n)$ matrices. In the Cartan
decomposition a choice of orthogonal basic Hamiltonians is used.
In the present decomposition we do not consider an orthogonal set
of basic Hamiltonians but the form of the Hamiltonians is imposed
by the physical system under consideration, fulfilling the
conditions of encoded universality \cite{DiVincpre01,DiVincNa00}

\begin{prop}\label{prop:SU8}
Let  $H_1, \, H_2,\, H_3,\, H_4, \, H_5$ are elements of the
algebra su$(2^3)$ such that the successive commutations of these
elements generate su$(2^3)$. Then  any element $U$ belonging to
the group SU$(4)$ is given by the relation:
\[
\begin{array}{rl}
  U=&{\rm  e}^{-i H_2 t_{63}}{\rm  e}^{-i H_1 t_{62}}
  \cdots {\rm  e}^{-i H_1
  t_6}\cdot\\
  &\cdot
  {\rm  e}^{-i H_5 t_5}{\rm  e}^{-i H_4 t_4}
  {\rm  e}^{-i H_3 t_3}{\rm  e}^{-i H_2 t_2}
  e^{-i H_1 t_1}
\end{array}
\]
\end{prop}

A similar Proposition can be stated in the case of the general
problem related to the  SU$(2^n)$ group. The above Proposition
concerns the problem of controllability of spin systems, in the
context of the Cartan decomposition technique. This problem was
solved \cite{KhaGla01} and other studies of the same problem by
different techniques have been recently proposed
\cite{ObRaWa99,AlDal01_Pre}.
 The general problem can
be formulated in a different way. From the theory of universal
gates \cite{DeuPRSL85} and the papers on the control of the
molecular systems \cite{Rama01,RaObSun00} it is well known that
the SU$(2^n)$ can be decomposed into simpler matrix factors with
SU$(2)$  and  SU$(4)$ structure. That means that the one and two
qubit gates are universal ones.  A systematic study of this
technique is under current investigation.

\end{appendix}


\begin{thebibliography}{99}


\bibitem{DeuPRSL85}
 D.  Deutsch, ``Quantum Theory, the Church-Turing principle and the
Universal Quantum Computer'', {Proc. Roy. Soc. London A}
\textbf{400}, 97-117 (1985).

\bibitem{BraLoFDP00}
  S.  Braunstein,   H.  -K.  Lo, Editors, ``Experimental proposals for
Quantum Computation'',    {Fort. Phys.} \textbf{48}, 765-1138
(2000).

\bibitem{RaViMoKo00}
  H.  Rabitz,   R.  de Vivie-Riedle,   M.   Motzkus,   K.  Kompa,
``Whither the future of controlling quantum phenomena?\,'',
   {Science} \textbf{288}, 824-828 (2000).

\bibitem{DiVincpre01}
  D.  P.  DiVincenzo,   D.  Bacon,   J.  Kempe,   D.  A.  Lidar,
  K.  B.  Whaley, ``Encoded Universality in physical implementations
of a Quantum Computer'', LANL e-print \textbf{quant-ph/0102140}.

\bibitem{DiVincNa00}
  D.  P.  DiVincenzo,   D.  Bacon,   J.  Kempe,   G.  Buckard,
  K.  B.  Whaley, 2000, ``Universal quantum computation with exchange
interaction'',    {Nature} (London) \textbf{408}, 339-342,
\textbf{quant-ph/0005116}.

\bibitem{JurSus75}
  V.  Jurdjevi\'{c},   H.  Sussmann, ``Control systems on Lie
groups'',    {J. Diff. Eq.} \textbf{12}, 313-329 (1975).

\bibitem{Rama95}
  V.  Ramakrishna,   M.  V.  Salapaka,   M.  Dahleh,   H.  Rabitz,
  A.  Peirce, ``Controllability of molecular systems'',    {Phys.
Rev. A} \textbf{51}, 960-966 (1995).

\bibitem{Rama01}
  V.  Ramakrishna, ``Control of molecular systems with very few
phases'',    {Chem. Phys.} \textbf{267}, 25-32 (2001)

\bibitem{KhaGla01}
  N.  Khaneja,   S.  J.  Glaser, ``Cartan decomposition of $Su(2^N)$
and control of spin systems'',    {Chem. Phys.} \textbf{267},
11-23 (2001), \textbf{quant-ph/0010100}.

\bibitem{FuSchSo01}
  H.  Fu,   S.  G.  Schirmer,   A.  I.  Solomon, ``Complete
Controllability of finite-level quantum systems'',    {J. Phys. A:
Math. and Gen.} \textbf{34}, 1679-1690 (2001),
\textbf{quant-ph/0102017}.

\bibitem{SchFuSo01}
  S.  G.  Schirmer,   H.  Fu,   A.  I.  Solomon, ``Complete
controllability of quantum systems'',    {Phys. Rev. A}
\textbf{63}, 025403 (2001), \textbf{quant-ph/0010031}.

\bibitem{NakaPRL97}
  Y.  Nakamura,   C.  D.  Chen,   J.  S.  Tsai, ``Spectroscopy of
energy-level splitting between two macroscopic quantum states of
charge coherently cuperposed by Josephson coupling'',    {Phys.
Rev. Lett.} \textbf{79}, 2328-2331 (1997).

\bibitem{AveSSC98}
  D.  V.  Averin, ``Adiabatic quantum computation with Cooper
pairs'',    {Sol. St. Comm.} \textbf{105}, 659-664 (1998),
\textbf{quant-ph/9706026}.

\bibitem{AveNa99}
  D.  V.  Averin, ``Solid-state qubits under control'',
  {Nature} (London)
\textbf{398}, 748-749 (1999).

\bibitem{ShnSchPRL97}
  A.  Shnirman,   G.  Sch\"on,   Z.  Hermon, ``Quantum manipulation of
small Josephson junctions'',    {Phys. Rev. Lett.} \textbf{79},
2371-2374 (1997), \textbf{quant-ph/9706016}.

\bibitem{MahkRMP00}
  Y.  Makhlin,   G.  Sch\"on,   A.  Shnirman, ``Quantum state
engineering with Josephson junction devices'',    {Rev. Mod.
Phys.} \textbf{73}, 357-400 (2000), \textbf{cond-mat/0011269}.

\bibitem{MakhNa99}
  Y.  Makhlin,   G.  Sch\"on,   A.  Shnirman, ``Josephson junction
qubits with controlled couplings'',    {Nature} (London)
\textbf{398}, 305-307 (1999), \textbf{cond-mat/9808067}.

\bibitem{Mahkpre99}
  Y.  Makhlin,   G.  Sch\"on,   A.  Shnirman, ``Josephson junction
qubits and the readout process by single electron Transistors'',
LANL e-print \textbf{cond-mat/9811029}.

\bibitem{NakaNa99}
  Y.  Nakamura,   Yu.  A.  Pashkin,   J.  S.  Tsai, ``Coherent control of
macroscopic quantum states in a single-Cooper-pair box'',
   {Nature} \textbf{398}, 786 (1999), \textbf{cond-mat/9904003}.

\bibitem{NakaJLTP00}
  Y.  Nakamura,   J.  S.  Tsai, ``Quantum-state control with a
single-Cooper-pair box'',    {J. Low Temp. Phys.} \textbf{118},
765-779 (2000).

\bibitem{BaBePRA95}
  A.  Barenco,   Ch.  H.  Bennett,   R.  Cleeve,   D.  P.  DiVincenzo,
  N.  Margolus,   P.  Shor,   T.  Sleator,   J.  A.  Smolin,
  H.  Weinfurter, ``Elementary gates for quantum computation'',
   {Phys. Rev. A} \textbf{52}, 3457-3467 (1995),
\textbf{quant-ph/9503016}.

\bibitem{Varadarajan}
  V.  S.  Varadarajan, \emph{Lie Groups, Lie Algebras and their
representations} ed. Prentice-Hall.

\bibitem{ObRaWa99}
  R.  J.  Ober,   V.  Ramakrishna,   E.  S.  Ward, ``On the role of
reachability and observability in NMR experimentation'',    {J.
Math. Chem.} \textbf{26}, 15-26 (1999).

\bibitem{AlDal01_Pre}
  F.  Albertini,   D.  D'Alessandro, ``The Lie algebra structure and
nonlinear controllability cf spin systems'', LANL e-print
\textbf{quant-ph/0106115}.

\bibitem{RaObSun00}
  V.  Ramakrishna,   R.  Ober,   X.  Sun, ``Explicit generation of
unitary transformations in a single atom or molecule''    {Phys.
Rev. A} \textbf{61}, 032106 (2000).

\end{thebibliography}
\end{document}